\documentclass[12pt,a4paper]{article}
\usepackage[utf8]{inputenc}
\usepackage{amsmath,amssymb}
\usepackage{graphicx}
\usepackage{hyperref}
\usepackage{float}
\usepackage{hyperref}
\usepackage{subcaption} 
\usepackage{float} 
\usepackage{caption}
\usepackage{subcaption}
\usepackage{physics}
\usepackage[sort]{cite}
\usepackage{siunitx}
\usepackage[margin=1in]{geometry}

\title{Spatially Mapping Phonon Drag in Ultrascaled 5-nm Silicon Nanowire Field-Effect Transistor Based on a Quantum Hydrodynamic Formalism}

\author{
Houssem Rezgui\textsuperscript{1,2*}, 
Giovanni Nastasi\textsuperscript{3}, 
Manuel Marcoux\textsuperscript{1}, 
Vittorio Romano\textsuperscript{4} \\
\small \textsuperscript{1}IMFT-CNRS, University of Toulouse, 7 Avenue du Colonel Roche, Toulouse, 31400, France \\
\small \textsuperscript{2}International Iberian Nanotechnology Laboratory (INL), Braga 4715-330, Portugal \\
\small \textsuperscript{3}University of Enna ``Kore'', Department of Engineering and Architecture, Enna, Italy \\
\small \textsuperscript{4}Department of Mathematics and Computer Science, University of Catania, Catania, Italy \\
\small \textsuperscript{*}Corresponding authors: 
\href{mailto:houssem.rezgui@toulouse-inp.fr}{houssem.rezgui@toulouse-inp.fr},}

\begin{document}

\maketitle

\begin{abstract}
The growing demand for better performance and lower thermal energy dissipation in nanoelectronic devices is the major driving force of the semiconductor industry's quest for future generations of nanotransistors. Over the past 15 years, the miniaturization of silicon-based nanoelectronics predicted by Moore's law has driven an aggressive scaling down of transistor structures, including materials, design, and geometries. In this regard, the electronic device community has expanded its focus to ultrascaled transistors targeting the 7-nm technology node and beyond. However, these emerging nanodevices also present thermal challenges that can limit carrier transport as a result of strong electron–phonon coupling. In this work, we investigate the physical origin of self-heating effects in an ultrascaled 5-nm silicon nanowire field-effect transistor. Based on a quantum hydrodynamic approach, we also provide an explanation of the phonon drag contribution to thermal conductivity. We report the impact of the phonon drag effect on the electrical and thermal performance of 5-nm gate-all-around silicon nanowire field-effect transistors. Our findings provide new insight into the origin of self-heating as a result of mutual electron–phonon coupling. Furthermore, we demonstrate that the phonon drag effect significantly reduces thermal conductivity by nearly 50\% under high-bias conditions.
\end{abstract}

\newpage

\section{Introduction}

Today, the investigation of nanoscale thermal engineering is regarded as an important prerequisite for thermoelectric devices \cite{1,2,3}, semiconductors \cite{4,5}, and emerging nanoelectronics \cite{6,7}. In particular, nanoelectronic devices have greatly benefited from the new development of complementary metal-oxide-semiconductor (CMOS) nanomanufacturing methods \cite{8,9,10}. Unlike classical planar MOS field effect transistors (MOSFET) which cannot scale down due to the device geometries and hot carrier degradation (HCD) \cite{11,12}. Although FinFETs are regularly used in many devices, such as static memory cells (SRAM) at the 14nm node, FinFET technology suffers from self-heating effects (SHEs) and thermal stress response \cite{13}. Therefore, novel scalable architectures have been explored for next-generation nanotransistors such as gate-all-around silicon (GAA) (Si) nanowire (NW) FETs \cite{14,15,16}. The GAA Si NWFETs are attractive for the electronic device community and the International Technology Roadmap for Semiconductor (ITRS) targets because of their excellent energy delay product (EDP) comparable to those of two-dimensional (2D) FETs \cite{17}. In addition, Si NWFETs have good electrostatic integrity and high-performance (HP) applications compatible with current 5nm CMOS devices \cite{17}. Following the ITRS 2026 edition, the Si NWFET will consider an ultra-short gate length $L_{g}$ of $5\,\mathrm{nm}$ with a diameter $\ d = 3\,\mathrm{nm}$ \cite{18}. Ultrascaled devices are expected to meet scaling challenges and continue Moore’s law trends; however, experimentally reporting the transfer characteristics of nanoscale transistors remains difficult due to their quantum mechanical nature, including effects such as ballistic transport, geometrical confinement, and carrier scattering.

In recent years, several suitable frameworks have been proposed to enhance the design performance and reliability of ultrascaled FETs because of the lack of experimental descriptions of ultrafast carrier transport in the nanoscale regime. The miniaturization of nanotransistors leads to an increase in the internal temperature higher than the operating temperatures of the device (that is, hot spot) \cite{8,19,20}. Theoretical simulations of ultrafast carrier dynamics are not a straightforward task, since the latter requires a full description of phonon--electron coupling \cite{21,22,23,24,25,26,27} and phonon-phonon scattering \cite{28,29,30}. Quantum ab initio methods based on first-principles calculations are very accurate in predicting the tunneling of electrons within the valence and conduction bands of the semiconductor material \cite{10,22}. Moreover, the non-equilibrium Green function (NEGF) approach, where the electron Green's function $G(E)$ is coupled to the phonon mode $D(w)$ through scattering self-energies, is efficient for considering electrothermal transport in nanoscale transistors \cite{23,25}. In addition, quantum hydrodynamic (QHD) models are derived to establish the relationship between the macroscopic quantity of the semiconductor material such as the electron mobility and the microscopic description (e.g., scattering events) \cite{31,32,33,34,35,36}. The quantum drift-diffusion model is obtained under the assumption that the mean free path (MFP) of the carrier is much larger than the device length. Using density gradient (DG) theory, a quantum drift-diffusion model is established in order to investigate the electron hydrodynamic transport \cite{33,36}.  DG theory describes the collective behavior of electron gas flow in a higher Knudsen regime analogous to the Mach number that manifests shock waves in fluid dynamics \cite{33}. In addition, the DG formalism takes into consideration quantum correction in terms of non-local strain gradient occurring due to the electron-lattice (i.e., phonon) interactions.

Hydrodynamics behavior is one of the most fruitful theoretical formalisms for describing emergent physical phenomena in many-body systems. Quantum hydrodynamic models have been employed to solve the electron--phonon (e-ph) Boltzmann transport equations (BTE) in low-dimensional materials \cite{31,32}. Macroscopic quantum modeling is computationally less expensive and accurate for realistic device physics. To meet situations in semiconductors, macroscopic description has been adapted through a similarity between Knudsen flow (i.e., the ratio between MFP and characteristic length) in gas dynamics and carriers (phonon and electron) transport in solids \cite{33}. The lowest order of the Chapman-Enskog expansion to the BTE yields the Navier-Stokes-Fourier equations for viscous fluids. For the slip regime ($Kn \approx 1$), the classical Fourier law fails due to the nonlocality of the temperature gradient which depends on the properties of the boundary \cite{37}. To describe gas flow in high Knudsen number, a four-moment non-equilibrium phonon distribution is obtained under the maximum entropy principle \cite{37}. The hydrodynamic formalism matches higher-order approximation moments where strong thermodynamic non-equilibrium effects occur due to the carrier scattering mechanism. For example, the Guyer-Krumhansl equation (GKE) was derived based on the maximum entropy method for the flow of phonon gas in dielectric crystals at low temperature \cite{38,39}. Figure 1 summarizes multiscale carrier transport from small Knudsen number to high-order Knudsen flow.


\begin{figure}[H]
    \centering
    \includegraphics[width=1\textwidth, keepaspectratio]{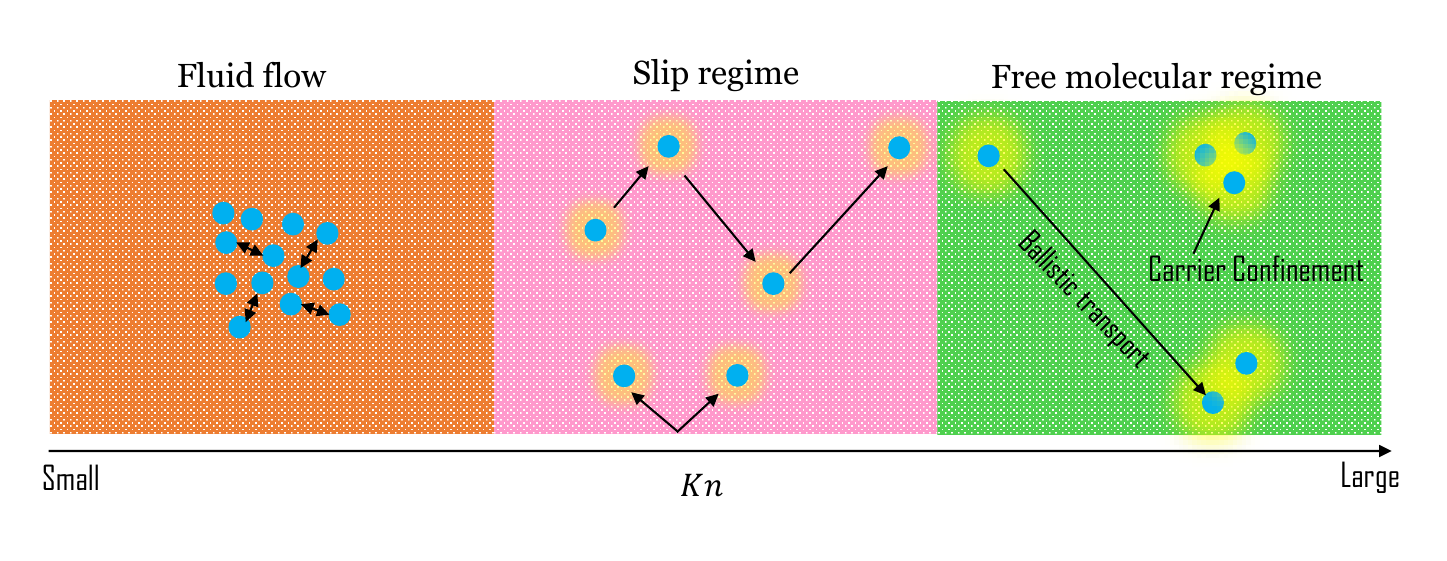}
    \vspace{-10pt} 
    \caption{Multiscale carrier transport in Knudsen regime.}
    \label{fig:fig1}
\end{figure}

Due to the presence of a high electric field, carrier degradation and self-heating effect strongly reduce the ON-state current of the ultrascaled device. Electron--phonon scattering plays an important role in affecting the performance of nanoscale CMOS devices \cite{10,21,22,23,24}. Electron--phonon coupling is a key factor for a deeper understanding of electrothermal transport \cite{23}. For a strong external electrical field, the non-equilibrium phonon distribution in the electron flow will provide momentum exchange through the channel carriers (Si NW), known as the phonon drag effect \cite{40}. Extensive efforts have been reported to address the effect of phonon drag in low-dimensional materials \cite{40,41,42,43,44,45}; however, in ultrascaled devices, phonon drag is still relatively unclear. The phonon drag effect was first recognized in Germanium and silicon to enhance the Seebeck coefficient at cryogenic temperatures \cite{45,46}. The phonon drag has recently been examined in GaN and AlN with a wide band gap into the nature of e-ph coupling and its impact on carrier mobility at room temperature \cite{40}. More recently, the influence of the drag effect on thermal conductivity has been investigated in 2D materials \cite{47}. The phonon drag effect has not received much attention in emerging nanotransistors because of less knowledge on the mutual interaction between out-of-equilibrium phonons and electrons through ultrascaled channel devices.

In this work, we investigate the phonon drag effect in 5-nm Si NW FET using the quantum hydrodynamic formalism. The present method will include two numerical schemes for electron and phonon transport. The aim of this paper is to combine both electron and phonon hydrodynamic transport into momentum balance equations. Furthermore, this computational framework will be able to capture the phonon drag impact on electrical transport in a heavily doped n-type 5nm GAA Si NW transistor. This work probes new insights into nanoscale electrothermal transport and into the nature of electron--phonon gas coupling. We organized this paper as follows: The hydrodynamic formalism is introduced in Sec. 2. First, we derive the electron hydrodynamic equation based on the DG theory. In addition, we provide a way to obtain the quantum-drift diffusion model. Furthermore, we derive the phonon hydrodynamic equation and the related boundary condition. In Sec. 3, we validate our proposed approach by comparing the effective thermal conductivity with theoretical and experimental results. We also compare the transfer characteristics $I_D$-$V_{GS}$ with the nonequilibrium Green function formalism. Moreover, we discuss the impact of the electron--phonon coupling on device performance. Finally, conclusions and remarks are presented in Sec. 4.

\section{Hydrodynamic formalism for nanoscale semiconductors}

Hydrodynamic transport is a branch of the science of macroscopic methods from equilibrium thermodynamic effect in fluid mechanics to extended irreversible thermodynamic in nanosystems \cite{48,49,50}. The phonon and electron hydrodynamic equations are quite similar to the phenomenological approach for rarefied gas in hypersonic flow at extremely high flight velocities (e.g., aircraft and guided missiles). The hydrodynamic characteristic of Bose-condensed gas is widely investigated in quantum shock waves \cite{51} and superfluid turbulence \cite{52}. Quite interestingly, the hydrodynamic behavior of electron and phonon transport in semiconductors exhibits characteristics similar to viscous fluids under a pressure gradient in the Poiseuille flow \cite{53,54}. Therefore, inspired by the kinetic theory of collective gas flow, we developed coupled e-ph hydrodynamic equations for nanoscale semiconductors.

\subsection{Electron hydrodynamic equation}

\subsubsection{Physical origin of macroscopic quantum transport in semiconductors}

The main feature of the hydrodynamic DG is to establish density gradient corrections to the electron gas flow. In other words, the DG enhances the drift-diffusion (DD) model by adding quantum contributions, and therefore the so-called quantum drift-diffusion model (QDD).  DG theory considers the lowest-order corrections of the quantum hydrodynamic effect, considering that the electron gas density depends on its non-local gradients \cite{33}. The inclusion of quantum nonlocality leads to a generalized quantum drift-diffusion current equation through a drag force exhibited by phonon scattering on the electron gas. In semiconductors, the momentum balance equations for electron gas are expressed as \cite{33,36}
\begin{equation}
\frac{\partial}{\partial t}\int_{V}^{}{n\ dV} + \int_{S}^{}{n \mathbf{u} \cdot \hat{\mathbf{n}} \, dS} = 0
\end{equation}
where $n$ is the electron density, $V$ is the control volume, $S$ is its surface, $\hat{\mathbf{n}}$ is the normal unit vector outward to $S$, and $\mathbf{u}$ is the electron velocity. Using DG theory, the energy balance equation for the electron gas is expressed as \cite{30}
\begin{equation}
\frac{\partial}{\partial t}\int_{V}^{}{m n \mathbf{u} \, dV} + \int_{S}^{}{m  n \left(\mathbf{u}\cdot \hat{\mathbf{n}}\right) \mathbf{u} \, dS} = - \int_{S}^{}{np^{n} \hat{\mathbf{n}} \, dS +}\int_{S}^{}{\sigma^{n} \hat{\mathbf{n}} \, dS + \int_{V}^{}{qn (\mathbf{E} + \mathbf{E}^{\mathbf{n}}) \, dV}}
\end{equation}
where $m$, $p^{n}$,$\sigma^{n}$ and $q$ are the mass, the pressure, the stress tensor and the charge of the electron, respectively. The term $qn\mathbf{E}$ represents the electrostatic force, and ${qn\mathbf{E}}^{\mathbf{n}}$ denotes the force exerted by the lattice vibrations (i.e. phonons) on the electron gas. Here, the electron--phonon coupling has two different contributions: (1) a dissipative part $,\ \mathbf{E}^{nd}$, known as the drag force, and (2) a non-dissipative part $\mathbf{E}^{nr},$ that refers to the carrier effective mass due to the lattice diffraction \cite{33}. The differential equation of the DG formalism is
\begin{equation}
m_{e}^{*}n\frac{d \mathbf{u}}{dt} = - \nabla p^{n} + \nabla \cdot \sigma^{n} + qn  (\mathbf{E} + \mathbf{E}^{nd})
\end{equation}
where $m_{e}^{*}$ is the effective mass of the electron. The usual form of the hydrodynamic DG equation can be written as follows.
\begin{equation}
m_{e}^{*}\frac{d\mathbf{u}}{dt} = - q\nabla\varphi_{DG}^{n} + {q\mathbf{E}}^{nd}
\end{equation}
where $\varphi_{DG}^{n}$ is the chemical electron potential defined as $\varphi_{DG}^{n} = \ \psi_{c} - \phi_{DG}^{n}$, in which $\phi_{DG}^{n}$ is the quasi-Fermi level and $q$ is the elementary charge. Here, $\psi_{c}$ denotes the classical potential that satisfies $\mathbf{E} = - \nabla\psi_{c}$. The principle of DG theory is to consider the strong electron--phonon coupling near the Fermi level. In nanoscale semiconductors, the DG theory aims to involve the physical origin of quantum behavior in electron gas flow. In the next section, we will derive the quantum DD model and related boundary conditions.

\subsubsection{Quantum drift-diffusion model}

In ultrascale devices, electrons are quantum mechanical objects that obey the effective-mass Schrödinger equation or the Hartree approach (“Schrödinger-Poisson”) \cite{55}. In the macroscopic description, the electron gas equations are derived from unified and clear mathematical descriptions based on the Boltzmann-Wigner transport equation (BWTE) \cite{31,56}. The Wigner formalism adequately accounts for the quantum transport of the carrier in the ultrascaled device through the Wigner function $f_{w}(t,\mathbf{x},\mathbf{k})$, which represents the quasi-distribution function of the electron. The BWTE is given by \cite{31,56,57}
\begin{equation}
\frac{\partial f_{w}(t,\mathbf{x},\mathbf{k})}{\partial t} + \frac{\hbar\mathbf{k}}{m_{e}^{*}}\cdot\nabla_{\mathbf{x}}f_{w}(t,\mathbf{x},\mathbf{k}) + \frac{q}{\hbar}\nabla_{\mathbf{x}}\psi\cdot\nabla_{\mathbf{k}}f_{w}(t,\mathbf{x},\mathbf{k}) = \mathcal{Q}\left( f_{w} \right) + \mathcal{C}\left( f_{w} \right)
\end{equation}
where $\mathbf{x} \in \Omega \subset \mathbb{R}^3$ is the position of the electron, $\mathbf{k} \in \mathcal{B} \subseteq \mathbb{R}^3$ is the vector of the electron wave with $\mathcal{B}$ being the first Brillouin zone, $\hbar$ is the reduced Planck constant, and $\psi$ is the electrostatic potential obeying the Poisson equation expressed as
\begin{equation}
\nabla_{\mathbf{x}}\cdot\left( \varepsilon\nabla_{\mathbf{x}}\psi(\mathbf{x}) \right) = q(N_{A} - N_{D} + n)
\end{equation}
where $\varepsilon$ is the dielectric constant, $N_{D}$ and $N_{A}$ are the doping concentrations of the donors and acceptors, respectively. Here, $n$ denotes the electron density associated with the Wigner function, $n = \int_{}^{}{f_{w}(t,\mathbf{x},\mathbf{k}) \, d\mathbf{k}}$. In equation (5), $\mathcal{C}\left( f_{w} \right)$ is often simplified with a semiclassical collisional operator taking into account scattering between electrons and phonons or impurities. It can be written as
\begin{equation}
\mathcal{C}\left( f_{w} \right) = \int_{}^{}{\left\lbrack S\left( \mathbf{k}',\mathbf{k} \right)f_{w}\left( t,\mathbf{x},\mathbf{k}' \right) - S\left( \mathbf{k},\mathbf{k}' \right)f_{w}(t,\mathbf{x},\mathbf{k}) \right\rbrack \, d\mathbf{k}'}
\end{equation}
where $S\left( \mathbf{k}',\mathbf{k} \right)$ is the transition rate, i.e. the probability distribution to have a scattering of an electron from the wave vector $\mathbf{k}'$ to $\mathbf{k}$. It can be determined in the specific case by means of Fermi's golden rule. $\mathcal{Q}\left( f_{w} \right)$ represents the quantum collisional operator defined as
\begin{equation}
\mathcal{Q}\left( f_{w} \right) = \int_{}^{}{V_{w}\left( \mathbf{x},\mathbf{k} - \mathbf{k}' \right)f_{w}\left( t,\mathbf{x},\mathbf{k}' \right) \, d\mathbf{k}'}
\end{equation}
where $V_{w}$ is the Wigner potential.
\begin{equation}
V_{w}(\mathbf{x},\mathbf{k}) = - q\frac{1}{i\hbar(2\pi)^{3}}\int_{}^{}{e^{- i \mathbf{k} \cdot \mathbf{x}'}\left\lbrack B\left( \mathbf{x} + \frac{\mathbf{x}'}{2} \right) - B\left( \mathbf{x} - \frac{\mathbf{x}'}{2} \right) \right\rbrack \, d\mathbf{x}'}
\end{equation}
with $B(\mathbf{x})$ being a function of the position modeling the potential barriers. In order to capture the quantum effect in ultrascaled devices, it is mandatory to modify the current continuity equation in the standard DD model. The implication of quantum potential correction allows an accurate interpretation of carrier tunneling, confinement, and strong carrier coupling effects. The current density which satisfies the QDD can be written as
\begin{equation}
j_{DG}^{n} = - D_{n}\nabla n + \mu_{n}n \nabla\psi_{c} + \mu_{n}n \nabla\psi_{q}
\end{equation}
where $\psi_{c}$ is the classical potential given by the sum of the solution of the Poisson equation (6) and the barrier potential $B$, while $\psi_{q}$ is the quantum potential given by Bohm's formalism \cite{58}
\begin{equation}
\psi_{q} = \frac{\hbar}{2qm_{e}^{*}r}\frac{\nabla_\mathbf{x}^{2}\left( \sqrt{n} \right)}{\sqrt{n}}.
\end{equation}

Equation (11) is the solution of the Schrödinger equation using the Wigner function. By substituting equation (11) into equation (10), we have
\begin{equation}
\mathbf{j}_{DG}^{n} =  - D_{n}\nabla n + \mu_{n}n\nabla\psi_{c} + {2\mu}_{n}n\nabla\left( b_{n}\frac{\nabla_\mathbf{x}^{2}\left( \sqrt{n} \right)}{\sqrt{n}} \right)
\end{equation}
where $b_{n}$ is defined as $b_{n} = \frac{\hbar^{2}}{4qm_{e}^{*}r_{n}}$ in which $\ r_{n}$ is a fitting parameter. For semiconductors, $r_{n}$ varies between 1 and 3 depending on the number of subband levels. Here, Equation (12) predicts the drift and diffusion of electron gas motion, also taking into account the quantum effects achieved by the second derivative of the electron concentration. It represents an approximation of the BWTE based on: (1) relaxation time approximation (RTA) and (2) the potential is slowly varying at any point in the ultrascaled channel device. In other words, the QDD model reflects the fast changes in the carrier dynamics with respect to the quantum hydrodynamic energy model. 

The QDD model assumes that the electrons have an equilibrium thermal state satisfying the Einstein relation, $\frac{D_{n}}{\mu_{n}} = \frac{k_{B}T}{q}$, where $k_{B}$ is the Boltzmann constant and T is the temperature distribution. Using the expression of the current density in terms of the gradient of the quasi-Fermi level, $\mathbf{j}_{DG}^{n} = \mu_{n}n\nabla\phi_{DG}^{n}$, equation (12) becomes
\begin{equation}
{\phi_{DG}^{n} - \psi}_{c} + \frac{k_{B}T}{q}\ln\left( \frac{n}{n_{0}} \right) = 2b_{n}\left( \frac{\nabla_\mathbf{x}^{2}\left( \sqrt{n} \right)}{\sqrt{n}} \right)
\end{equation}
where $n_{0}$ is the intrinsic carrier (i.e., low doping concentration). Equation (13) can be expressed as
\begin{equation}
{\phi_{DG}^{n} = \psi}_{c} - \varphi_{DD}^{n} + 2b_{n}\left( \frac{\nabla_\mathbf{x}^{2}\left( \sqrt{n} \right)}{\sqrt{n}} \right)
\end{equation}
where $\varphi_{DD}^{n}\ $ is the chemical potential given by the classical DD model defined as $\varphi_{DD}^{n} = \frac{k_{B}T}{q}\ln\left( \frac{n}{n_{0}} \right)$. In order to establish a full quantum drift-diffusion model, one requires appropriate boundary conditions at the interface between the semiconductor and the oxide. Here we assume that the contact is ohmic, whereas the normal component of the quantum force should vanish along the boundary. Therefore, Equation (11) gives the following
\begin{equation}
\hat{\mathbf{n}} \cdot \left( b_{n}\nabla_\mathbf{x}\sqrt{n} \right) = 0.
\end{equation}
For the DG, the quantum potential can be written as
\begin{equation}
\nabla \cdot \left( b_{n}\nabla_\mathbf{x}\sqrt{n} \right) - \frac{\sqrt{n}}{2}\psi_{q} = 0.
\end{equation}
Although the insulating boundary condition is very useful at the interface, it overestimates the QC effects at the oxide layer. In addition, the electron-wave function can penetrate the oxide region. For this reason, the following boundary conditions are prescribed  \cite{59}:
\begin{equation}
\hat{\mathbf{n}} \cdot \left( b_{n\,OX}\nabla_\mathbf{x}\sqrt{n} \right) = \frac{b_{n\,OX}}{x_{np}}\sqrt{n_{0}}
\end{equation}
where $b_{n\,OX}$ is the linear gradient parameter for the oxide $b_{n\,OX} = \frac{\hslash^{2}}{4qm_{n\,OX}^{*}r_{n}}$, in which $m_{n\,OX}^{*}$ is the electron effective mass in the oxide layer and $x_{np}$ is the penetration depth obtained from the Wentzel-Kramers-Brillouin (WKB) approximation \cite{59}:
\begin{equation}
x_{np} = \frac{\hbar}{\sqrt{2qm_{n\,OX}^{*}\Phi_{B}}}
\end{equation}
where $\Phi_{B}$ is the potential barrier height. Here, we wish to demonstrate an efficient and new methodology to tackle ultrascaled device modeling challenges, including phonon-electron coupling and non-equilibrium thermodynamic effects in sufficiently high electric fields and lattice temperatures.

\subsection{Multiscale thermal transport in ultrascaled transistors}

\subsubsection{Phonon hydrodynamic equation}

To achieve a full hydrodynamic window, the phonon distribution functions should include high-order gradients of the heat flux. Therefore, in this case, the phonon hydrodynamic equation (PHE) is considered for non-equilibrium thermal transport. Similarly to gas flow, the behavior of the phonon is predicted by ph-BTE \cite{29,37,60}
\begin{equation}
\frac{\partial g}{\partial t} + \mathbf{v}\cdot \nabla_{\mathbf{x}} g = C(g)
\end{equation}
where $g(t,\mathbf{x},\mathbf{k}_p)$ is the phonon distribution function, $\mathbf{x}$ is the position vector, $\mathbf{k}_p$ is the phonon wave vector, $\mathbf{v}$ is the velocity of the phonon group, and $C(g)$ is the scattering term that includes the normal ($N$) and resistive ($R$) phonon scattering processes. In general, equation (19) is set for each phonon mode (i.e. optical, acoustic, etc.). At cryogenic temperature, phonons are quantum bosons that satisfy the Bose-Einstein statistics with total momentum conservation during normal phonon-phonon scattering (i.e., $N$ dominates the $R$ process) \cite{38}. However, at room temperature, the momentum is not conserved as a result of Umklapp scattering. It is very difficult to evaluate the full scattering term $C(g)$ due to the complex nature of phonon transport in extremely low-dimensional materials. In addition, hydrodynamic heat transport in semiconductors is a common belief in non-Fourier thermal conduction at room temperature \cite{37}. In general, non-Fourier heat transfer requires isotropic distribution and a gray-Debye approximation where identical branches of acoustic phonons are treated with a constant relaxation time \cite{37}. 
Consequently, the single-mode relaxation time approximation (SMRTA) is assumed to solve the phonon BTE \cite{37,60,61,62}
\begin{equation}
\frac{\partial g}{\partial t} + \mathbf{v}\cdot \nabla_{\mathbf{x}}g = - \frac{g - g_{0}}{\tau_{R}}\ 
\end{equation}
where $g_{0}$ is the equilibrium distribution function given by the Planck distribution and $\tau_{R}\ $ is the relaxation time of resistive phonon-phonon scattering. In a mathematical point of view, the PHE has a similar GK-equation derived based on ph-BTE. In another world, the PHE is exactly the GK-equation, but derived at room temperature where R dominates the N process \cite{37}. The PHE is a macroscopic hydrodynamic model based on field variables obtained by the integration over the wave vector. In this case, the heat flux, $\mathbf{Q}$ is then expressed as
\begin{equation}
\mathbf{Q} = \int \mathbf{v} \hbar\omega \,g d\mathbf{k}_p
\end{equation}
where $\omega = \omega\left( \mathbf{k}_{p} \right)\ $ is the dispersion relation of phonons. The high-order approximation of $\mathbf{Q}$ is derived in terms of nonlocal terms of the gradient of the heat flux. In phonon kinetic theory, the PHE gives \cite{37}
\begin{equation}
\mathbf{Q} + \tau_{R}\frac{\partial \mathbf{Q}}{\partial t} = - \kappa\nabla T + \frac{l^{2}}{5}(\frac{1}{3}\nabla\nabla\cdot\mathbf{Q} + \nabla^{2}\mathbf{Q})
\end{equation}
where $\kappa$ is the bulk thermal conductivity and $l$ is the phonon MFP defined as $l = v\tau_{R}$. Equation (22) represents a hydrodynamic formalism that can investigate multiscale thermal transport in semiconductors. Here, the PHE includes a second-order derivative, $\nabla\nabla\cdot\mathbf{Q}$, which constitutes the non-equilibrium phonon gas flow. To properly establish non-Fourier effects in low-dimensional materials, an effective thermal conductivity concept should be carefully considered in multiscale hydrodynamic heat transport.

\subsubsection{Effective thermal conductivity}

Gate-all-around architecture has successfully created new perspectives in semiconductor nanowires and state-of-the-art nanoscale FETs. The manufacturing of ultrascaled FETs is still a technology under investigation and requires additional engineering innovation. Following the newly updated International Roadmap for Devices and Systems (IRDS) 2028 \cite{63}, the structure of the Si GAA NWFET consists of an ultra-short channel of $L_{g} = 5\,\mathrm{nm}$ as shown in Figure 2. 


\begin{figure}[H]
    \centering
    \includegraphics[width=0.7\textwidth, keepaspectratio]{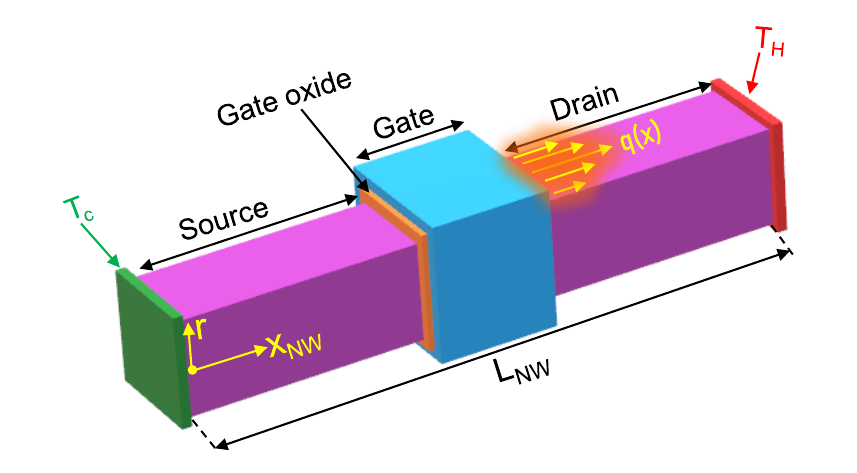}
    \vspace{-10pt} 
    \caption{Schematic representation of ultrafast thermal transport in nanoscale FETs. Here, $x_{\mathrm{NW}}$ denotes the longitudinal coordinate of the nanowire axis.}
    \label{fig:fig2}
\end{figure}

In the nanoscale framework, phonons are considered primary heat carriers in semiconductors and metals. A fundamental understanding of nanoscale heat transfer including semiconductors \cite{64,65,66,67}, porous materials \cite{68}, interfacial thermal resistance \cite{69,70,71}, and nanodevices \cite{72,73,74} is necessary to prevent thermal challenges in existing and future nanoengineering. In early-stage heating (picosecond time scale), both non-equilibrium thermal transport and quantum phonon confinement resulting in hotspots are supposed to be dominant physical mechanisms. For ultrascaled devices (sub-10 nm nanostructures), heat carriers exhibit a large mean free path that causes ballistic thermal conduction \cite{75}. Strong ballistic transport will therefore affect the thermal conductivity because of the phonon-boundary scattering and more pronounced phonon-drag effects. In this case, it is evident to develop an efficient approach that can be useful for size-dependent thermal conductivity. For steady-state heat flow through the channel, Equation (22) gives
\begin{equation}
Q_{x}(r) = - \kappa\frac{dT}{dx} + \frac{l^{2}}{5}\frac{d^{2}Q_{x}(r)}{dr^{2}} + \frac{l^{2}}{5}\frac{1}{r}\frac{dQ_{x}(r)}{dr}.
\end{equation}
where $r$ is the radius of the nanowire. The size-dependent effective thermal conductivity, $\kappa_{eff}$, is obtained from equation (23)
\begin{equation}
\kappa_{eff}\  = \frac{\int_{}^{}{Q_{x}(r)d\Gamma}}{\frac{S\Delta T}{L}}
\end{equation}
where $\Delta T$ is the temperature difference, $L$ is the length of the nanowire, and $S$ is the surface area. For a small Knudsen number and fully dominant resistive scattering ($\frac{1}{\tau_{R}} \rightarrow \infty$), Equation (22) recovers the classical Fourier's law (i.e., $\mathbf{Q} = - \kappa\nabla T$) which describes a fully diffusive thermal transport. In order to invoke interfacial heat transfer across the oxide and the ultrascaled channel, a well-known slip/jump boundary condition is needed to capture non-equilibrium effects caused by the interaction between the phonon gas flow and the wall within the Knudsen layer. The heat flux slip boundary is a Boltzmann transport solution via the Chapman-Enskog expansion. Thus, the slip boundary condition is therefore obtained from the tangential heat flux near the surface \cite{29,37,70}:
\begin{equation}
Q_{t} = C_{W}{l}\frac{dQ_{t}}{dn}
\end{equation}
where $C_{W}$ is a positive parameter that indicates a reduction in heat flux due to phonon scattering at surface roughness and $n$ is the surface unit normal vector. Similarly to the velocity slip condition, $C_{W}$ is very similar to the accommodation coefficient associated with the interaction of the rarefied gas near the solid wall \cite{29,70}. The jump-type boundary condition is commonly used in modeling strong non-equilibrium heat transfer at high Knudsen numbers and in predicting the Kapitza resistance between two materials. For high-temperature distributions, diffuse phonon-boundary scattering ($C_{W} = 1$) is included to properly account for nonlocal effects.

\subsection{Electron--phonon coupling: Quantum hydrodynamic formalism}

In modern nanoelectronics, phonon--electron coupling is vital to understanding the underlying physical phenomena responsible for the undesirable thermal degradation. In this section, we present a coupled hydrodynamic DG with phonon hydrodynamic equation including the interaction of the electron--phonon gas. The balance equation for energy conservation can be written as \cite{76}
\begin{equation}
C\frac{dT}{dt} + \nabla\cdot\mathbf{Q} = H
\end{equation}
where $C$ is the heat capacity and $H$ is the term of heat generation due to the electron--phonon coupling. Here, we present an enhanced electrothermal model taking into account electron--phonon scattering and quantum potential contribution. The source term $H$ is expressed as
\begin{equation}
H = \mathbf{j}_{DG}^{n}\cdot \mathbf{E} + (R - G)(E_{g} + 3k_{B}T)
\end{equation}
where $(R - G)$ is the rate of electron-hole recombination generation and $E_{g}$ is the energy band gap of the semiconductor. We developed an electrothermal model capable of combining both electron and phonon gas transport in ultrascaled channel transistors as shown in Figure 3(a). Our proposed quantum hydrodynamic formalism provides deep insight in ultrascale device modeling under (1) high-field conditions, (2) strong electron--phonon coupling, and (3) phonon drag effect. The hydrodynamic DG and the PHE are solved using a finite element method (FEM) with extremely fine mesh. Here, we present a novel formalism based on the hydrodynamic semiconductor equations for theoretical modeling of ultrafast carrier transport in ultrascaled FETs. Figure 3(b) illustrates the electron and phonon tracing scheme based on finite element analysis. A trace FEM is developed to simulate the trajectories of phonons and electrons under partial differential equations (PDEs). The FEM is considerably less computationally expensive and therefore could be employed in investigating multiscale multiphysics problems in complex geometries. The framework contains two main models: (1) the hydrodynamic DG equation for electron transport through the nanodevice and (2) the phonon hydrodynamic equation for ultrafast non-Fourier thermal conduction. We present a novel multiphysics coupling of nanoscale electrothermal transport in an extremely small system.


\begin{figure}[H]
    \centering
    \includegraphics[width=1\textwidth, keepaspectratio]{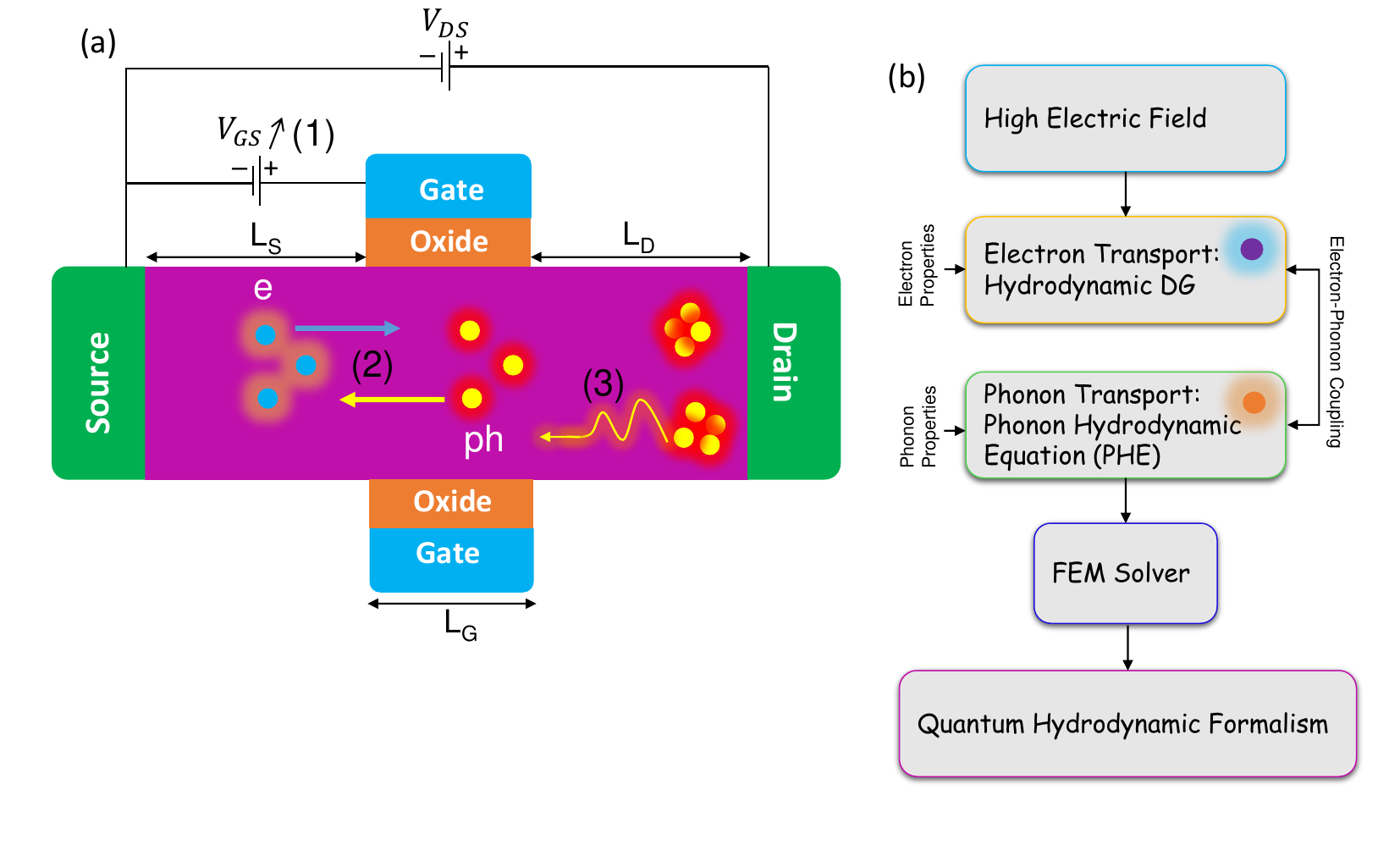}
    \vspace{-10pt} 
    \caption{(a) Electron--phonon (e-ph) interaction in ultrascaled gate-all around field-effect transistor. (b) Numerical scheme for quantum hydrodynamic formalism.}
    \label{fig:fig3}
\end{figure}

To investigate a realistic thermal transport, the phonon properties ($\kappa$, $c$ and ${l}$) are ab initio coefficients. For electron transport through the n-type $5\,\mathrm{nm}$ GAA Si NWFET, the hydrodynamic DG is solved for ${r}_{n} = 3$. The phonon and electron properties are reported in Tables 1 and 2, respectively. Additionally, in our QHD approach, phonons and electrons are addressed under steady-state conditions.

\begin{table}[htbp]
\centering
\caption{Ab initio inputs for phonon properties in silicon at different temperatures. The ab initio coefficients $\kappa$, $l$ and $C$ are obtained from \cite{28} and \cite{30}.}
\label{tab:phonon_properties}
\resizebox{\textwidth}{!}{%
\begin{tabular}{|c|c|c|c|c|c|c|c|c|c|c|}
\hline
Temperature & 300 K & 320 K & 340 K & 360 K & 380 K & 400 K & 420 K & 440 K & 460 K & 480 K \\
\hline
$\l$ (nm) & 185 & 165 & 155 & 139 & 133 & 128 & 110 & 107 & 98 & 93 \\
\hline
$\kappa$ (W$\cdot$m$^{-1}$$\cdot$K$^{-1}$) & 145 & 130.6 & 121.8 & 112.1 & 105.7 & 103 & 94.2 & 92 & 87.2 & 85 \\
\hline
$C$ (MJ$\cdot$m$^{-3}$$\cdot$K$^{-1}$) & 1.63 & 1.671 & 1.7 & 1.73 & 1.762 & 1.786 & 1.8 & 1.824 & 1.84 & 1.854 \\
\hline
\end{tabular}%
}
\end{table}

\begin{table}[htbp]
\centering
\caption{Values of key parameters for electron transport.}
\label{tab:electron_transport}
\begin{tabular}{|c|c|}
\hline
\textbf{Parameter} & \textbf{Value} \\
\hline
$N_{D}$ & $1 \times 10^{20}\ \mathrm{cm}^{-3}$ \\
\hline
$m_{e}^{*}$ & $0.28\, m_{e}$ \\
\hline
$m_{nox}^{*}$ & $0.14\, m_{e}$ \\
\hline
$x_{np}$ & $2.95\ \text{\AA}$ \\
\hline
$r_{n}$ & 3 \\
\hline
$\Phi_{B}$ & $3.1\ \mathrm{eV}$ \\
\hline
\end{tabular}
\end{table}

\section{Results and discussion}

\subsection{Model validation}

The quantum hydrodynamic formalism is an approach that describes the collective motion of carriers in a material, similar to how fluid dynamics describes the flow of liquids or gases. In this framework, phonons, which are typically considered individual scattering particles, are instead treated collectively, leading to fluid-like behavior. Acoustic phonons are low-energy states that correspond to the coherent motion of carriers, and they are mainly responsible for heat transfer at the nanoscale. In the concept of nanoscale thermal transport in materials, particularly at the microscopic and atomic levels, phonons play a key role in predicting effective thermal conductivity (ETC) \cite{77,78}. In the first stage, the ETC of the silicon thin films is compared with the theoretical model (ab initio method \cite{64}), the empirical model \cite{65} and the experimental measurements \cite{66,67} as shown in Figure 4.


\begin{figure}[htbp]
    \centering
    \includegraphics[width=\textwidth, keepaspectratio]{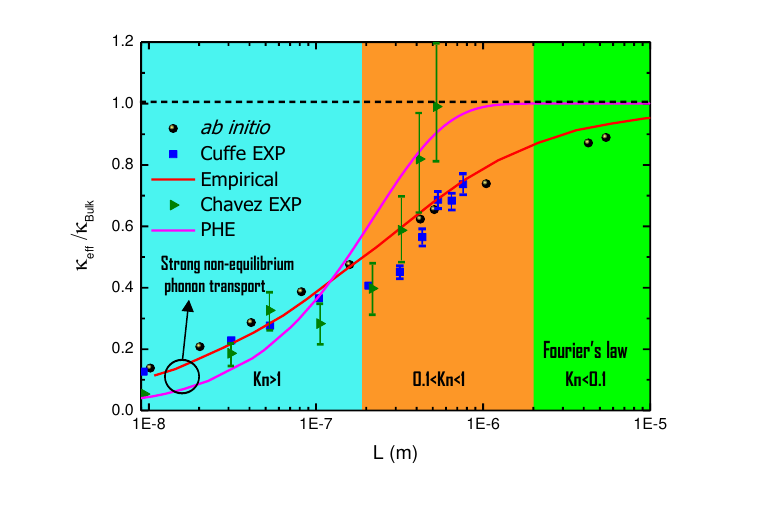}
    \caption{Comparison of normalized thermal conductivity of silicon nanofilms as a function of the system length ($L$, defined as the distance between the hot and cold sides). The dashed line indicates the diffusive limit, $\kappa_{\text{eff}}/\kappa_{\text{Bulk}} = 1$. The plot also includes previously reported size-dependent thermal conductivities from theoretical models and experiments for comparison.}
    \label{fig:fig4}
\end{figure}

The PHE provides reliable predictions even for extremely small structures and demonstrates clear advantages in capturing sub-continuum features of multiscale heat transport, spanning from the ballistic regime ($Kn > 1$) to the diffusive regime ($Kn \ll 1$). The results confirm that the thermal conductivity exhibits a strong dependence on the Knudsen number: at high $Kn$, non-equilibrium phonon transport dominates, and the associated nonlocal effects substantially degrade the effective thermal conductivity of silicon. In the intermediate Knudsen regime ($0.1 < Kn < 1$), a noticeable discrepancy is observed between the experimental datasets of Cuffe \textit{et al.}~\cite{66} and Chávez-Angel \textit{et al.}~\cite{67}. Cuffe \textit{et al.}, using non-contact transient thermal grating measurements, reported a reduction in conductivity from about $74\%$ to $13\%$ of the bulk value, while Chávez-Angel \textit{et al.}, employing Raman thermometry on ultra-thin high-quality Si membranes, systematically measured lower conductivities for the thinnest films. This divergence can be attributed to differences in sample quality, surface conditions (roughness and boundary scattering), and measurement techniques, which strongly affect phonon boundary scattering and transport pathways.

\begin{figure}[!ht]
    \centering
    \includegraphics[width=0.9\textwidth]{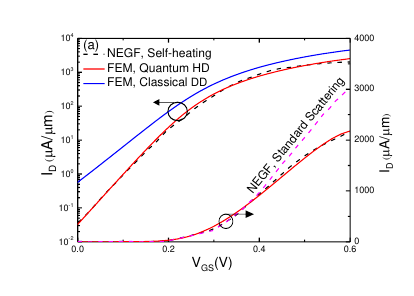}\\[-2pt] 
    \includegraphics[width=0.9\textwidth]{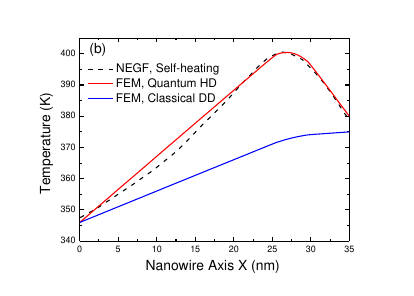}
    \caption{(a) Transfer characteristics at $V_{DS}=0.6$\,V for the ultrascaled Si GAA NWFET with HfO$_2$ gate oxide. Short dashed lines denote the NEGF with standard scattering (i.e., without self-heating effects) and nanoscale transistors \cite{24}. (b) Temperature distribution along the nanowire axis at $V_{GS}=0.4$\,V and $V_{DS}=0.6$\,V. Here, the classical DD is coupled with Fourier's law. Data from \cite{24} are only included for comparison.}
    \label{fig:fig5}
\end{figure}

Figure 5(a) plots the transfer characteristics $I_{D} - V_{GS}$ of the ultrascaled $5\,\mathrm{nm}$ Si GAA NWFET at $V_{DS} = 0.6\,\mathrm{V}$. In this simulation, the Si NW is surrounded by an oxide layer with $t_{ox} = 3\,\mathrm{nm}$ of HfO$_2$ ($\varepsilon = 20$). The Si NW channel has a diameter of 3 nm and the source/drain extensions of $L_{S} = L_{D} = 15\,\mathrm{nm}$. The doping concentration is fixed at $N_{D} = 1 \times 10^{20}\,\mathrm{cm}^{-3}$,  which was chosen to reflect the high doping levels typically employed in ultrascaled  Si nanowire transistors in order to ensure strong carrier injection and to suppress short-channel effects \cite{24}. To clarify and briefly validate the profile $I_{D} - V_{GS}$, our results are compared with two NEGF approaches \cite{24}: (1) NEGF with self-heating effects: assumes that electrons interact with the non-equilibrium phonon population, and (2) NEGF with standard scattering: supposes that electrons scatter with equilibrium phonons at a constant temperature, $T = 300\,\mathrm{K}$. From a device modeling point of view, the NEGF including the nonequilibrium phonon distribution is more reasonable because phonons exhibit a nonlocal behavior at the nanoscale. Our proposed QHD formalism agrees well with the data obtained by the NEGF approach, including self-heating effects. However, the classical DD model fails to reproduce the NEGF results because it neglects the electron--phonon interaction and quantum effects. The drift-diffusion model overestimates the increase in current flow, while the QHD ratio provides a more consistent prediction. For phonon-electron coupling and when the phonon drag effect is considered, the electron will transfer its momentum to the phonons, creating a non-equilibrium phonon gas distribution which reduces the electrical mobility and therefore affects the drain current. Figure 5(b) shows the temperature distribution along the Si NW for $V_{GS} = 0.4\,\mathrm{V}$ and $V_{DS} = 0.6\,\mathrm{V}$. Note that the classical DD is obtained by coupling the conventional drift-diffusion model with the classical Fourier's law. Obviously, the classical DD formalism is not able to predict the response to temperature overshooting. The spatial position of this peak is located near the drain side of the gate, where the energy transfer from electrons to phonons is maximized due to high carrier velocity and strong phonon scattering. This results in localized phonon heating,  which manifests as the observed temperature overshoot, highlighting the crucial role of phonon drag in limiting carrier mobility and affecting the drain current.

\begin{table}[htbp]
\centering
\caption{Summary of heat and electrical transport based on the classical and quantum hydrodynamic approaches.}
\label{tab:transport_approaches}
\begin{tabular}{|p{2.5cm}|p{5.5cm}|p{5.5cm}|}
\hline
\textbf{Approach} & \textbf{Heat transport equations} & \textbf{Electrical transport equations} \\
\hline
Classical & $\mathbf{Q} = - \kappa \nabla T$ & $\mathbf{j}_{DD}^{n} = - D_{n} \nabla n + \mu_{n} n \nabla \psi_{c}$ \\
\hline
Quantum hydrodynamic & $\begin{aligned}&\mathbf{Q} + \tau_{R} \frac{\partial \mathbf{Q}}{\partial t} = - \kappa \nabla T \\ &+ \frac{l^{2}}{5} \left( \frac{1}{3} \nabla\nabla{\cdot}\mathbf{Q} + \nabla^{2} \mathbf{Q} \right)\end{aligned}$ & $\begin{aligned}&\mathbf{j}_{DG}^{n} = - D_{n} \nabla n + \mu_{n} n \nabla \psi_{c} \\ & + 2\mu_{n} n \nabla \left( b_{n} \frac{\nabla_{\mathbf{x}}^{2} \sqrt{n}}{\sqrt{n}} \right)\end{aligned}$ \\
\hline
\end{tabular}
\end{table}

In order to compare the classical DD and quantum DD formalisms, Table 3 summarizes the main differences between the semiconductor transport equations. Because the channel length is very short, the electron--phonon interaction will significantly degrade the lattice temperature, causing a self-heating state. Self-heating is expected to drastically reduce the drain current under high bias conditions (i.e., high ${\ V}_{GS}$). In other words, the self-heating phenomenon originates from the phonon drag effect when electron--phonon coupling becomes dominant in ultrashort channel devices.

\begin{figure}[H]
    \vspace{-10pt} 
    \centering
    \includegraphics[width=0.9\textwidth, keepaspectratio]{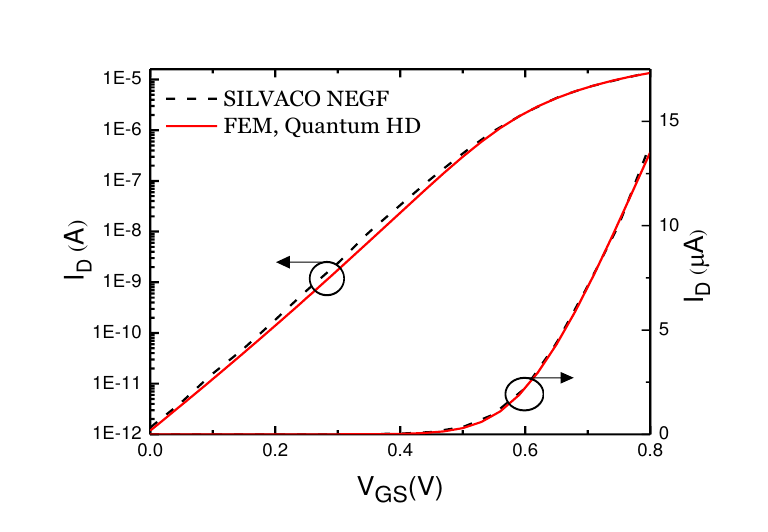} 
    \vspace{-5pt} 
    \caption{Transfer characteristics for the ultrascaled Si GAA NWFET with SiO$_2$ gate oxide. The Silvaco NEGF results are performed based on the non-equilibrium Green's function under the effective mass approximation coupled with the Poisson's equation \cite{78}. Data from \cite{78} are only included for comparison.}
    \vspace{-10pt} 
    \label{fig:fig6}
\end{figure}

To ensure an efficient QHD approach, we compare our results with the NEGF SILVACO simulations. The curve $I_{D} - V_{GS}$ is obtained using the QHD scheme in 2 hours using an Intel Xeon CPU E5-2650 (64 Go memory RAM) with a $V_{GS}$ step of 0.025\,V. The DG is written in MATLAB and combined with COMSOL multiphysics based on the Livelink interface \cite{79}. We focus on the output characteristic of the Si NWFET with the SiO$_2$ oxide layer of thickness $,\ t_{ox} = 0.5\,\mathrm{nm}$. The work function of the gate contact is about $4.05\ \mathrm{eV}$ and the dielectric constant is $\varepsilon = 3.9$. Figure 6 shows the transfer characteristics $I_{D} - V_{GS}$ of the Si GAA NWFET at $V_{DS} = 0.6\,\mathrm{V}$ simulated by the NEGF SILVACO model \cite{80} and our QHD formalism. Note that QHD allows quantum tunneling effects through a specific boundary condition (Equation 14). Moreover, ballistic current transport is involved in our multiphysics modeling of ultrascaled nanodevices. By simultaneously solving the PHE and the DG, the effect of a non-equilibrium phonon could be captured using the QHD formalism. In the next section, we will demonstrate the impact of electron--phonon coupling on conduction-band edges.

\subsection{Electrothermal effects: electron--phonon coupling}

In the classical drift-diffusion approach, carriers are in the thermal equilibrium regime where Fourier's law is still applicable. In ultrascaled transistors, quantum ballistic transport governs the carrier dynamics within the device. Here, the electrothermal effect refers to the physical mechanism in which electrons gain energy from the field and lose it by interacting with phonons or interfaces. For high bias, the electric field generates a nonequilibrium distribution in a confined geometry (5nm-channel region). The momentum exchange between extra electrons and nonequilibrium phonons is responsible for the emergence of the drag effect. We note again that the phonon drag effect requires a non-equilibrium state near the conduction-band edges.

\begin{figure}[H]
    \vspace{-10pt} 
    \centering
    \includegraphics[width=0.9\textwidth, keepaspectratio]{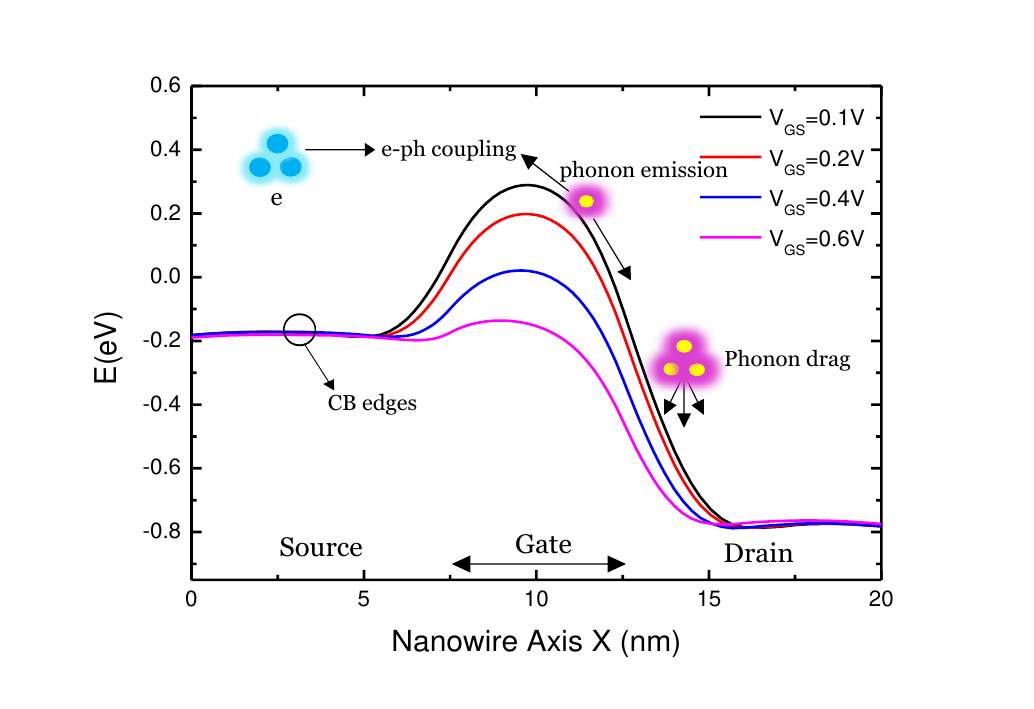} 
    \vspace{-5pt} 
    \caption{Conduction-band edges along the transport direction at $V_{DS} = 0.6\,\mathrm{V}$ for the Si GAA NWFET with SiO$_2$ gate oxide. The drag effect is a consequence of heavy population of nonequilibrium phonons close to the drain region. Phonon drag effect might change the electron direction (back-scattering mechanism) and limit electrons from reaching the drain contact.}
    \vspace{-10pt} 
    \label{fig:fig7}
\end{figure}

Figure 7 plots the edges of the conduction band (CB) in the nanowire at $V_{DS} = 0.6\,\mathrm{V}$ for different $V_{GS}$. Increasing $\ V_{GS}$, the potential decreases in the channel region. The phonon emission will oblige the electron to occupy a lower-energy state below the conduction band. Along the channel direction, the electron--phonon coupling follows the potential energy. At high bias conditions $\ (V_{GS} = 0.6\,\mathrm{V})$, the e-ph scattering mechanism allows strong non-equilibrium phonons due to the large applied electric field. In this case, the electron will be significantly dragged by the nonequilibrium phonon at the drain contact end of the channel region. We believe that the drag effect is caused by the non-equilibrium phonon under a non-local temperature gradient. The phonon drag gradually degrades the electron mobility as a result of the insufficient momentum close to the drain region because of weak electron velocity. For low $V_{GS} (0.1\,\mathrm{V}$ and $0.2\,\mathrm{V})$, the potential energy remains stable because the drag effect is found to be negligible (absence of dissipative force, $E^{nd} \approx 0$). The phonon-drag phenomena arise naturally when the e-ph coupling takes place; however, the drag will affect the device performance under high bias conditions.

\begin{figure}[H]
    \centering
    \includegraphics[width=1\textwidth, keepaspectratio]{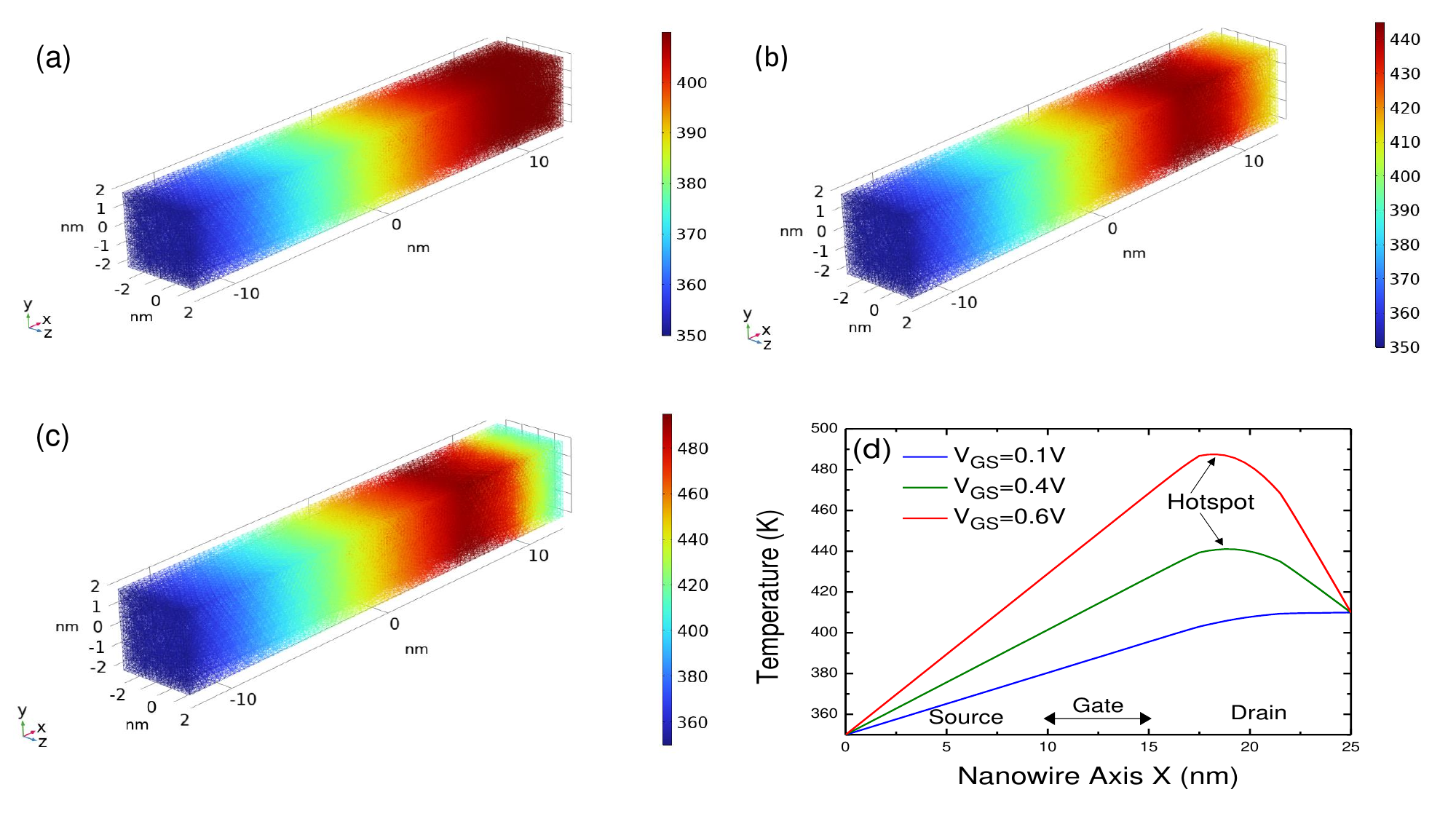}
    \includegraphics[width=0.85\textwidth, keepaspectratio]{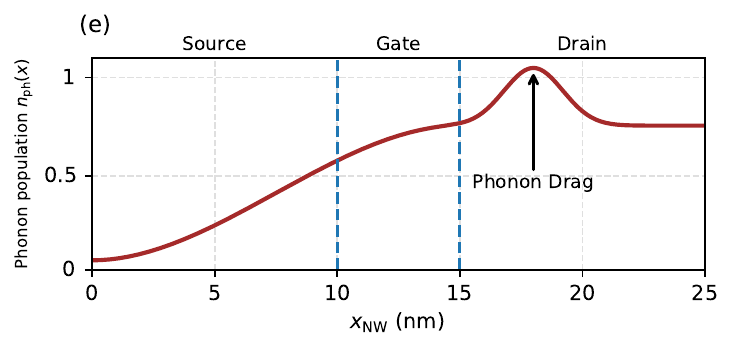}
    \vspace{-10pt} 
    \caption{Temperature distribution at \( V_{DS} = 0.6 \, \text{V} \) in the 5nm Si GAA NWFET:  (a) \( V_{GS} = 0.1 \, \text{V} \),  (b) \( V_{GS} = 0.4 \, \text{V} \),  (c) \( V_{GS} = 0.6 \, \text{V} \), and  (d) Temperature profiles along the nanowire for different \( V_{GS} \), (e) Phonon population $n_{\mathrm{ph}}(x)$ along the Si GAA NWFET at \( V_{GS} = 0.6 \, \text{V} \).}
    \label{fig:fig8}
\end{figure}

Next, we focus on evaluating the overall thermal performance of the state-of-the-art $5\,\mathrm{nm}$ Si GAA NWFET. Figure 8 shows the temperature distribution in the nanotransistor for different bias conditions. At $V_{GS} = 0.1\,\mathrm{V}$, the temperature gradually increases until it reaches a maximum near the drain contact, as shown in Figure 8(a). However, for $V_{GS} = 0.4$ and $0.6\,\mathrm{V}$, the temperature is overshooting and causing a hot-spot area. When the applied electrical current increases, the device temperature increases noticeably and exhibits important peaks, as shown in Figures 8(b) and (c). The peaks correspond to an accumulated non-equilibrium phonon population because of the strong coupling. In addition, the phonon drag effect will tend to generate an additional amount of temperature because of the backscattering process. In this case, the self-heating phenomenon originates from the fact that the phonon drag directly contributes to the increasing lattice temperature. In Figure 8(d), the temperature peaks in the drain can be clearly identified. Furthermore, it is demonstrated that for reasonably high bias conditions, the power dissipated by the non-equilibrium phonon population leads to a significant amount of the internal temperature in terms of self-heating effects. In addition, its spatial distributions prove that the creation of a hot-spot temperature is evidently associated with the non-equilibrium phonon state (i.e., the phonon drag effect). Figure~8(e) presents the normalized phonon population $n_{\mathrm{ph}}(x)$ as a function of the nanowire coordinate $x_{\mathrm{NW}}$ at $V_{GS}=0.6\,\mathrm{V}$. The phonon population remains low in the source region, increases within the channel due to electron–phonon interactions, and reaches a pronounced maximum near the drain side. 
This peak evidences the phonon drag induced by momentum transfer from high-velocity carriers to the lattice, directly supporting our interpretation of the temperature overshoot reported in Figs.~8(b)–(d).

In thermoelectric applications, the phonon drag effect can potentially boost the figure-of-merit ZT by increasing the Seebeck coefficient through an extra current induced via electron--phonon coupling. Therefore, phonon drag contributes to the improvement of energy harvesting in thermoelectric systems that convert thermal energy into electrical current. In contrast, for ultrascaled transistors, the increase of the current density will degrade the thermal conductivity. More particularly, the drag between an electron and a non-equilibrium phonon could lead to significant quantum confinement and, therefore, to a severe decrease of the thermal conduction feature.

\begin{figure}[H]
    \vspace{-10pt} 
    \centering
    \includegraphics[width=0.9\textwidth, keepaspectratio]{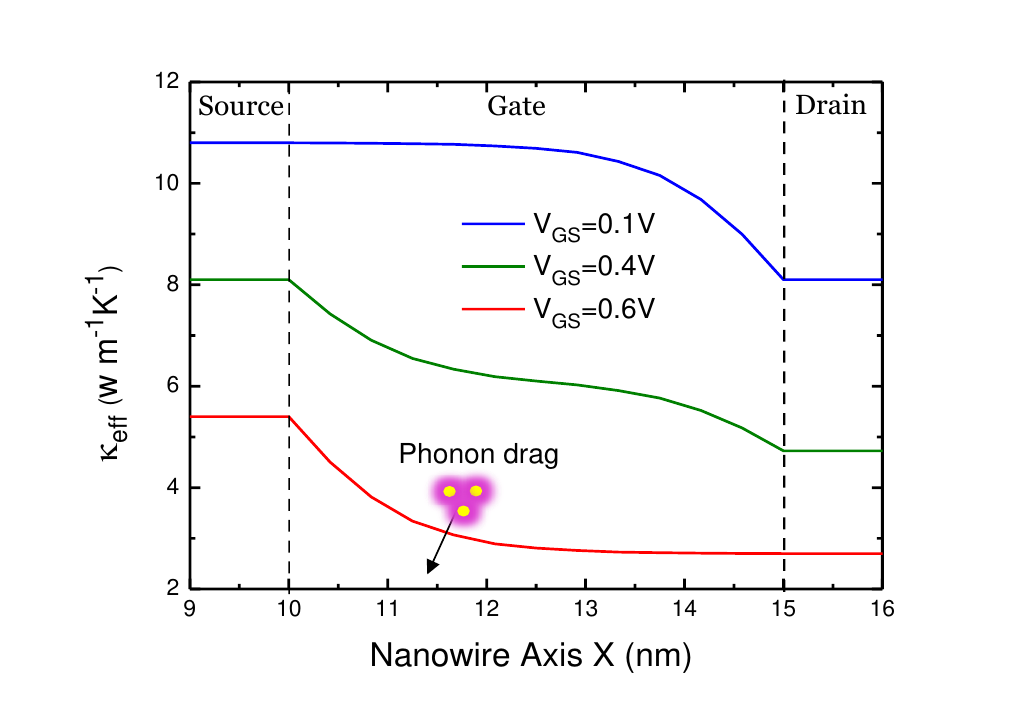} 
    \vspace{-5pt} 
    \caption{The extracted effective thermal conductivity along the channel region for different bias conditions.}
    \vspace{-10pt} 
    \label{fig:fig9}
\end{figure}

Figure 9 shows the thermal conductivity degradation along the channel region for different bias conditions. The thermal conductivity decreases with increasing gate voltage (i.e., high electron current). The high current density in ultrascaled Si NWFET generates a significant phonon out-of-equilibrium distribution. The cause of the poor thermal conductivity is the change in phonon energy states. More specifically, the phonon drag effect (out-of-equilibrium phonon state) makes it harder to evacuate the dissipated heat in the channel region and causes a reduced heat-carrying capacity. It is also found that the phonon drag affects not only the thermal conduction close to the drain contact, but also inside the channel region. We estimate that the drag effect can significantly reduce the thermal conductivity by nearly 23\% and 50\% under the bias condition of $V_{GS} = 0.4$ and $0.6\,\mathrm{V}$, respectively.

\begin{figure}[H]
    \setlength{\textfloatsep}{5pt}  
    \setlength{\floatsep}{5pt}      
    \setlength{\intextsep}{5pt}     
    
    \vspace*{-10pt} 
    \centering
    \includegraphics[width=0.9\textwidth, keepaspectratio]{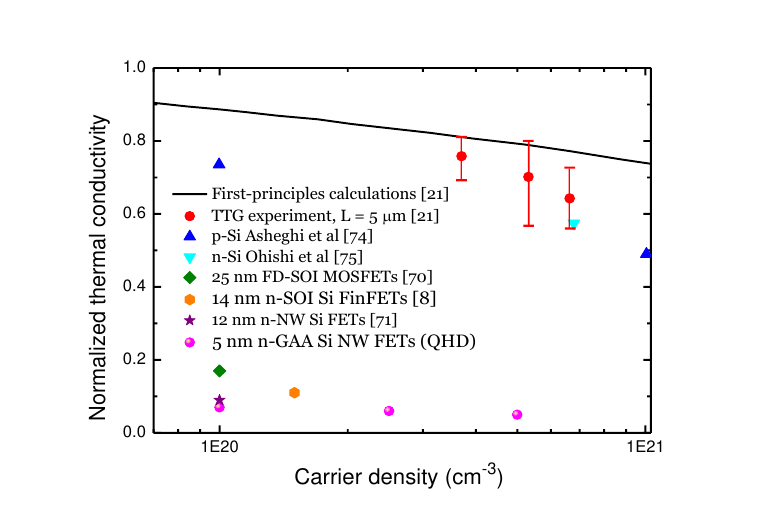} 
    \vspace{-5pt} 
    \caption{Normalized thermal conductivity as a function of the carrier density for bulk silicon and nanoscale Si MOSFETs. We note that phonon-boundary scattering is considered for all MOSFET technologies. For the ultrascaled 5 nm GAA Si NWFET, both phonon-boundary scattering and electron--phonon coupling are accounted.}
    \vspace{-15pt} 
    \label{fig:fig10}
\end{figure}

Figure 10 shows the normalized thermal conductivities as a function of the high carrier density for nanoscale transistors \cite{8,73,74} and bulk samples \cite{21,77,78}. Here, both the transient thermal grating (TTG) experiment (with grating period $L = 5\,\mu\mathrm{m}$) and the first-principles calculations involve phonon-electron coupling \cite{21}. These results uncover the impact of the electron--phonon interaction on the heat transfer in heavily doped semiconductors. It is found that strong scattering of phonons with electrons plays a major role in the reduction of the thermal conductivity. Doping is particularly significant in low-dimensional materials, where it controls carrier transport and conductivities. For emerging nanoscale transistors, one would expect weak thermal conductivity. Fully depleted (FD) MOSFETs have better thermal conductivity because they are less sensitive to boundary scattering and size effects. The 10 nm and 5 nm technology nodes NWFETs have lower thermal conductivities below 12 W\,m$^{-1}$\,K$^{-1}$.

\subsection{Device characteristics}

As modern nanotransistors continue to scale down, phonon-electron coupling will further reduce the ON-state current and pose additional thermal instability. The capability of manipulating the electron--phonon coupling is a key factor in the study of electrothermal behavior in emerging nanoscale CMOS technologies. Fig.~7 describes the impact of the e-ph coupling on the edges of the conduction band. The strong coupling effect will produce a nonequilibrium phonon population responsible for the desirable drag effect. To avoid an important phonon drag effect, nanoscale transistors should be operated in low bias conditions (less than 0.6\,V). Furthermore, the weak thermal conductivities and increased hot-spot temperatures in ultrascaled FETs indicate that thermal cooling strategies will be a critical challenge for next-generation nanotransistors. As discussed in Fig.~8 and 9, the thermal transport properties in the short channel can change significantly by varying the gate voltage. At $V_{GS} = 0.6\,\mathrm{V}$, due to the maximum generation of phonons, a hot spot temperature (480\,K) appears, indicating the presence of self-heating where the phonon-drag effect is predominant. The phonon-drag effect modifies the electrothermal characteristics of the device when the electron is traveling ballistically from source to drain and interacts with the non-equilibrium phonon population. Hence, it is challenging to enhance the thermal performance of ultrascaled FETs, but it is possible to reduce the frictional heat production by manipulating non-equilibrium phonon states and the motion of interacting surfaces.

\section{Conclusion}

In summary, we have inquired about the electrical and thermal performance of ultrascaled 5-nm Si NWFETs based on a quantum hydrodynamic approach. This work aims to explain several phenomenological processes in the phonon and electron hydrodynamic transport during ultrafast physical response. Our study provides an accurate framework for the investigation of phonon-drag and self-heating effects in ultrascaled nanodevices. We have argued the drag behavior using an advanced quantum drift-diffusion model coupled with a phonon hydrodynamic equation that includes mutual electron--phonon gas interaction. We demonstrate that a combination of density gradient theory and a non-Fourier heat model is very useful for describing the strong electron--phonon coupling in heavily doped nanodevices under high bias conditions. This formalism allows us to spatially map the phonon-drag effect along the nanowire. The drag effect has been found to contribute to an increase in lattice temperature through a high non-equilibrium phonon population at the end of the gate close to the drain region (${12\ \mathrm{nm} < x}_{NW} < 20\ \mathrm{nm}$). In high electrical fields and confined geometries, the electron--phonon scenario will lead to a nonequilibrium situation in which electrons are dragged by non-equilibrium phonons.

The redistribution of the non-equilibrium phonon population is expected to be produced by the sliding motion and the nanofriction response \cite{81}. In addition, it is known in general that frictional heat dissipation occurs because of the shear stress of the adjacent layers. In future work, we will consider nanoscale thermal friction in ultrascale nanodevices.

\section*{Data availability statement}

All data supporting the findings of this study are available upon reasonable request from the authors.

\section*{Acknowledgments}

The authors thank Dr. Albert Beardo for providing the ab initio coefficients for phonon transport in silicon. G.N. and V.R. acknowledge the financial support from MUR progetto PRIN “Transport phonema in low dimensional structures: models, simulations and theoretical aspects” CUP E53D23005900006.

\section*{Credit author statement}

\textbf{Houssem Rezgui:} Investigations; Format analysis; Writing - Original Draft; Methodology, Software; Visualization.
\textbf{Giovanni Nastasi:} Investigations; Writing - Original Draft; Analysis; Visualization.
\textbf{Manuel Marcoux:} Investigations; Writing - Original Draft; Analysis.
\textbf{Vittorio Romano:} Funding acquisition; Validation, Writing - Review \& Editing.

\end{document}